\documentclass[aps,prl,superscriptaddress,reprint]{revtex4-1}

\usepackage{graphicx}
\usepackage{amsmath}
\usepackage{amssymb}
\usepackage{dsfont}

\newcommand{\kb}[0]{k_{\rm B}}

\begin{document}

\title{The Gibbs paradox revisited from the fluctuation theorem with absolute irreversibility}

\author{Y\^uto Murashita}
\affiliation{Department of Physics, University of Tokyo, 7-3-1 Hongo, Bunkyo-ku, Tokyo, 113-0033, Japan}
\author{Masahito Ueda}
\affiliation{Department of Physics, University of Tokyo, 7-3-1 Hongo, Bunkyo-ku, Tokyo, 113-0033, Japan}
\affiliation{RIKEN Center for Emergent Matter Science (CEMS), Wako, Saitama 351-0198, Japan}
\date{\today}

\begin{abstract}
The inclusion of the factor $\ln (1/N!)$ in the thermodynamic entropy proposed by Gibbs is shown to be equivalent to the validity of the fluctuation theorem with absolute irreversibility for gas mixing.
\end{abstract}

\maketitle

\begin{figure}[b]
	\includegraphics[width=0.45\textwidth]{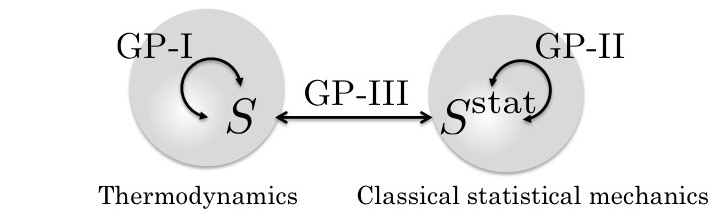}
	\caption{\label{GPQR}
		Three aspects of the Gibbs paradox.
		GP-I, GP-II and GP-III respectively concern the consistency within thermodynamics, the consistency within classical statistical mechanics, and the consistency  between thermodynamics and classical statistical mechanics.
	}
\end{figure}

As analyzed by van Kampen \cite{Kampen1984}, what is collectively called the Gibbs paradox actually involves the following three distinct issues (see Fig.~\ref{GPQR}):
\begin{description}
\setlength{\itemsep}{0cm}
\item[{\normalfont GP-I}] consistency within thermodynamics \cite{Gibbs1875,Kampen1984,Jaynes1992},
\item[{\normalfont GP-II}] consistency within classical statistical mechanics \cite{Ehrenfest1921,Kampen1984},
\item[{\normalfont GP-III}] consistency between thermodynamics and classical statistical mechanics \cite{Gibbs1902,Kampen1984,Jaynes1992}.
\end{description}
As detailed below, all of them have been settled in the thermodynamic limit \cite{Gibbs1875,Gibbs1902,Ehrenfest1921,Kampen1984,Jaynes1992}.
However, in view of the growing interest in small thermodynamic systems  \cite{SekimotoBook,JarzynskiBook,SeifertRev}, it is worth revisiting the Gibbs paradox in this context.
The conventional resolutions for GP-I \cite{Kampen1984,Jaynes1992} and GP-II \cite{Kampen1984} apply to small thermodynamic systems as well.
However, the one for GP-III cannot be applied to small thermodynamic systems \cite{Jaynes1992}.

The essence of GP-III is to determine ambiguity function $f(N)$ which gives the difference between the thermodynamic entropy $S$ and the statistical-mechanical entropy $S^{\rm stat}$ as
\begin{equation}\label{Deff}
	S(T,V,N)
	=
	S^{\rm stat} (T,V,N) + \kb f(N).
\end{equation}
In the thermodynamic limit, extensivity of the thermodynamic entropy leads to $f(N)=-N\ln N+cN$, where $c$ is a constant \cite{Pauli1973,Jaynes1992}.
However, extensivity breaks down in small thermodynamic systems.
In this Letter, we  show that in small thermodynamic systems the fluctuation theorem \cite{EvansCohenMorriss1993, GallavottiCohen1995PRL, Jarzynski1997PRL, Crooks1999, HatanoSasa2001, Seifert2005,JarzynskiRev,SeifertRev,BroeckEspositoRev,JarzynskiBook} with absolute irreversibility \cite{MurashitaFunoUeda2014} in place of extensivity leads to $f(N)=\ln(1/N!)+cN$.

{\it A brief history.}
In the 1870s, Gibbs put forth the problem of GP-I \cite[pp.227-229]{Gibbs1875}.
Whereas identical-gas mixing at the same temperature and pressure does not produce an extensive amount of entropy, different-gas mixing does.
The fact that the latter entropy production stays constant even when the difference between two species becomes infinitesimal seems paradoxical.
Gibbs ascribe this to how we define thermodynamic states.
This idea was further elaborated later.
For example, van Kampen \cite[Sec.2-4]{Kampen1984} argued that
the Clausius equality defines only the defference in entropy of a system where the particle number $N$ is conserved.
Therefore, to determine the $N$-dependence, we have to introduce a reversible process in which $N$ is varied.
For identical gases, opening a channel between the two partitions is reversible.
For different gases, this process is irreversible and consequently we have to invoke the familiar reversible procedure with semi-permeable walls.
In this way, difference between identical- and different-gas mixings reduces the two different processes used to extend the definition of entropy.
Which of the two processes we should use depends on whether we intend to use the semi-permeable walls or on how we operationally define the identical thermodynamic states.
``Thus, whether such a process is reversible or not depends on how discriminating the observer is. The expression for the entropy [...] depends on whether or not he is able and willing to distinguish between the molecules A and B''~\cite[pp.306-307]{Kampen1984}.
As Jaynes put it \cite[Sec.5]{Jaynes1992},
the thermodynamic entropy has ``anthropomorphic'' nature in that its value depends on ``human information.''

About a quarter century later, Gibbs discussed GP-III in his renowned textbook \cite[Chap.XV]{Gibbs1902} by introducing two types of phase spaces, ``the generic phase'' and ``the specific phase.''
In ``the generic phase,'' a set of points in phase space are identified when any point in the set can be obtained by permutation of particles of other points,
whereas in ``the specific phase,'' they are not.
Gibbs preferred ``the generic phase'' to ``the specific phase'' because the former results in no entropy production upon identical-gas mixing \cite[pp.206-207]{Gibbs1902}.
Thus, Gibbs introduced the factor $\ln (1/N!)$ in classical statistical mechanics to ensure the consistency with thermodynamic observation.
Van Kampen \cite[Sec.~9]{Kampen1984} later claimed that this prescription is guaranteed by convention (namely, by assuming $f(N)=0$ in Eq.~(\ref{Deff}) for the classical statistical-mechanical entropy in the generic phase).
However, Jaynes \cite[Sec.9]{Jaynes1992} was able to determine $f(N)$ from the extensivity of the thermodynamic entropy as we see later.

In the 1910s and 1920s, the inclusion of the factor $\ln(1/N!)$ was discussed within statistical mechanics (GP-II).
In particular, Eherenfest and Trkal considered the association and dissociation of molecules from a viewpoint of atomic phase space \cite{Ehrenfest1921},
and concluded that the factor $\ln(1/N!)$ is needed in calculating the chemical potential to guarantee the consistency within classical statistical mechanics.
Later, van Kampen \cite[Sec.7]{Kampen1984} derived the same factor based on combinatorics by considering a system connected to an infinitely large particle reservoir.

Although the Gibbs paradox can be understood within classical theory as described above,
Planck subsequently connected it with quantum distinguishability (e.g. see Ref.~\cite{Planck1925}).
Since then, the view that quantum theory resolves the Gibbs paradox has prevailed among physicists; for example, Schr\"odinger stated, ``The modern view [i.e. quantum theory] solves this paradox by declaring that in the second case [i.e. identical-gas mixing] there is no real diffusion, because exchange between like particles is not a real event [...]'' \cite[pp.61]{Schrodinger1946}. 
Now, in many standard textbooks (for example, see Refs.~\cite{Feynman1972, Zubarev1974, LandauLifshitz1975, Kubo1978, Callen1985, Prigogine1998}), the factor $\ln(1/N!)$ is attributed to the quantum indistinguishability of identical particles.
However, as van Kampen \cite{Kampen1984} and Jaynes \cite{Jaynes1992} pointed out, the quantum resolution is in fact irrelevant to the Gibbs paradox.
In fact, the quantum resolution suffers two major difficulties  \cite{Kampen1984, Jaynes1992}.
First, it cannot apply to classical mesoscopic systems such as a colloidal system.
Colloidal particles are not quantum-mechanically identical due to their distinguishable internal degrees of freedom.
Therefore, the quantum resolution fails to explain the factor $\ln(1/N!)$, although it is needed to explain experimental results (e.g. see Refs.~\cite{Frenkel2014, CatesManoharan2015}).
More importantly, quantum statistical mechanics is at the same position as classical mechanics with respect to Eq.~(\ref{Deff}).
In other words, the quantum statistical-mechanical entropy $S^{\rm q\mathchar`-stat}$ satisfies $S=S^{\rm q\mathchar`-stat}+f^{\rm q}(N)$.
Therefore, we must go through with the procedure to determine the ambiguity function $f^{\rm q}(N)$ as we do in the classical case.
To refer to this fact, van Kampen \cite[pp.311]{Kampen1984} catchily states, ``the Gibbs paradox is no different in quantum mechanics, it is only less manifest.''

{\it Conventional resolution to GP-III.}
GP-III concerns the relation between the thermodynamic entropy $S$ and the classical statistical-mechanical entropy $S^{\rm stat}$.
While the latter is defined as the Gibbs entropy, namely the Shannon entropy of the canonical state,
the former is defined through the Clausius equality as $\Delta S = \int_{\rm q.s.}\delta Q/T$, where $\delta Q$ is the heat transferred from the heat reservoir with temperature $T$ to the system and the integration is taken along a quasi-static process.
We note that the particle number $N$ cannot be varied in this process; if we start with open systems, the Clausius definition should be modified \cite[Sec.9]{Jaynes1992} (see {\it Discussions} for detail).
Therefore, the thermodynamic entropy defined by the Clausius equality has ambiguity about $N$.
Consequently, the relation between the thermodynamic entropy and statistical-mechanical entropy should involve some ambiguity function $f(N)$ as in Eq.~(\ref{Deff}) \cite[Eq.~(14)]{Kampen1984} \cite[Eq.~(16)]{Jaynes1992}.
The theme of GP-III is to demonstrate $f(N)=\ln (1/N!)$ (up to a constant per particle) for the classical statistical-mechanical entropy in the specific phase (or equivalently $f(N)=0$ for the one in the generic phase).
Incidentally, we note that ``exactly the same argument will apply in quantum statistical mechanics'' \cite[Sec.9]{Jaynes1992}.
Therefore, GP-III is not the issue of classical statistical mechanics in particular, but any statistical mechanics in general.

Jaynes \cite[Sec.9]{Jaynes1992} resolved this issue for free particles by analogy with the way by which Pauli phenomenologically determined the $N$-dependence of the thermodynamic entropy \cite[pp.38-39]{Pauli1973}.
By requiring that the thermodynamic entropy satisfy extensivity
\begin{equation}\label{Ext}
	S(T,qV,qN)=qS(T,V,N),\ ^\forall q>0,
\end{equation}
and by invoking the formula of the statistical-mechanical entropy for an ideal gas
$
	S^{\rm stat}(T,V,N)
	=
	N\kb ((3/2)\ln T + \ln V +{\rm const.}),
$
one can show 
\begin{equation}\label{NlnN}
	f(N)=Nf(1)-N\ln N.
\end{equation}
Here, the first term on the right-hand side of Eq.~(\ref{NlnN}) represents an intrinsic entropy per particle, while the second term amounts to the factor $-\ln N!$ in the large-$N$ limit.
Thus, the requirement of extensivity leads to the desired factor in the thermodynamic limit ($N\to\infty$).

However, extensivity, in general, breaks down in small thermodynamic systems, and therefore the Pauli-Jaynes resolution cannot apply to them.
Jaynes \cite[Sec.9]{Jaynes1992} expressed his concern about this point by stating, ``The Pauli correction was an important step in the direction of getting ``the bulk of things'' right pragmatically; but it ignores the small deviations from extensivity that are essential for treatment of some effects; and in any event it is not a fundamental theoretical principle. A truly general and quantitatively accurate definition of entropy must appeal to a deeper principle which is hardly recognized in the current literature, [...]''
Thus, we need a different guiding principle to determine $f(N)$ in small thermodynamic systems.

{\it Method.}
We use the fluctuation theorem \cite{EvansCohenMorriss1993, GallavottiCohen1995PRL, Jarzynski1997PRL, Crooks1999, HatanoSasa2001, Seifert2005,JarzynskiRev,SeifertRev,BroeckEspositoRev} as the guiding principle for small thermodynamic systems.
The fluctuation theorem, especially the Jarzynski equality $\langle e^{-\beta(W-\Delta F)} \rangle = 1$ \cite{Jarzynski1997PRL}, determines the equilibrium free-energy difference $\Delta F$ between two configurations from statistics of work $W$ in a nonequilibrium process.
Unfortunately, the Jarzynski equality cannot apply to gas mixing because the initial state is not a global canonical equilibrium state of the entire box, which invalidates the prerequisite of the Jarzynski equality.

We overcome this difficulty by applying the fluctuation theorem in the presence of absolute irreversibility 
$
	\langle e^{-\beta(W-\Delta F)} \rangle
	=
	1-\lambda
$
\cite{MurashitaFunoUeda2014}
which allows an initial canonical state with spatial constraints.
Here, the term $\lambda$ is the absolutely irreversible probability which is mathematically defined as the integral of the singular part of the time-reversed probability measure with respect to the forward probability measure \cite{MurashitaFunoUeda2014}.
For this singular part, the Crooks fluctuation theorem \cite{Crooks1999} is ill-defined because the dissipative work is divergent with singularly strong irreversibility.
This is why the Jarzynski equality \cite{Jarzynski1997PRL} should be modified in absolutely irreversible processes.
We can calculate $\lambda$ as the sum of the probabilities of the time-reversed events whose corresponding forward events vanish.

\begin{figure}
	\includegraphics[width=0.40\textwidth]{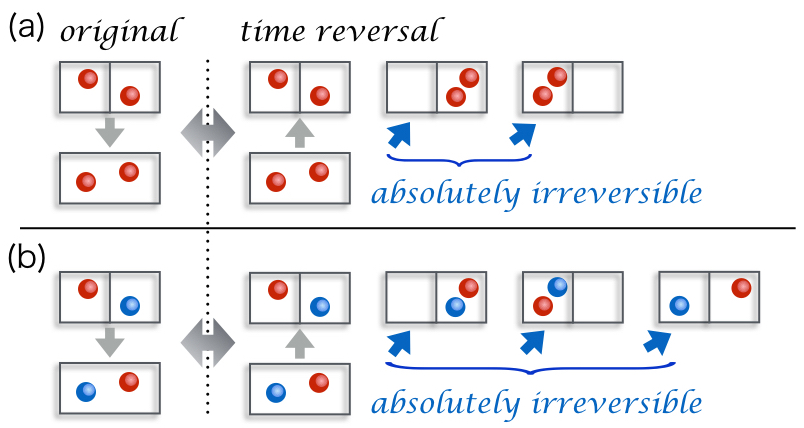}
	\caption{\label{Fig:GM}
		Time reversal and absolute irreversibility.
		The time-reversed process of gas mixing is the wall insertion for a global thermal equilibrium state.
		(a) Original and time-reversed processes of mixing of two identical gases.
		In the time-reversed process, the rightmost two events in which the state does not return to the original one have no counterparts in the forward process.
		Therefore, they are absolutely irreversible.
		(b) Original and time-reversed processes of mixing of two different gases.
	The probability of absolutely irreversible events is larger due to the presence of the rightmost event which has no counterpart in (a).
	}
\end{figure}

To demonstrate the existence of absolute irreversibility in gas mixing, we consider the time-reversed process of gas mixing, namely, wall insertion as illustrated in Fig.~\ref{Fig:GM}.
The events indicated by blue arrows are absolutely irreversible because they have no counterparts in the forward process.
The absolutely irreversible probability $\lambda$ is given by the sum of the probabilities of these events.
Thus, absolute irreversibility of gas mixing originates from fluctuations in the number of particles in each box after wall insertion
and may be interpreted as an inevitable loss of information on the particle number in each box before gas mixing \cite{AshidaFunoMurashitaUeda2014}.

{\it Main result.}
We now show that the validity of the fluctuation theorem with absolute irreversibility 
\begin{equation}\label{FTAI}
	\langle e^{-\beta(W-\Delta F)} \rangle
	=
	1-\lambda
\end{equation}
for gas mixing in the isothermal environment is equivalent to the following choice of the ambiguity function:
\begin{equation}\label{N!}
	f(N)
	=
	Nf(1)-\ln N!.
\end{equation}
In particular, if the free-energy difference is calculated from Eq.~(\ref{FTAI}), the thermodynamic entropy is defined and $f(N)$ is given by Eq.~(\ref{N!}).
Recall that, in the thermodynamic limit, the validity of extensivity (\ref{Ext}) is equivalent to Eq.~(\ref{NlnN}).
Thus, the fluctuation theorem with absolute irreversibility~(\ref{FTAI}) in small thermodynamic systems takes the place of extensivity (\ref{Ext}) in the thermodynamic limit.
Note that Eq.~(\ref{N!}) differs from Eq.~(\ref{NlnN}) at the subleading order, namely in the mesoscopic regime.

{\it Proof.}
First, we assume Eq.~(\ref{FTAI}) to derive Eq.~(\ref{N!}).
We consider mixing of identical gases and that of different gases as illustrated in Fig.~\ref{Fig:GM},
and assume that gases are initially in thermal equilibrium with inverse temperature $\beta$.
We assume that the gas particles have the same mass and the interaction potential between two different particles is the same as that between two identical particles.
Therefore the Hamiltonians for the two cases are identical.
We also assume that the interactions do not break additivity.
By the term additivity, we mean that the thermodynamic entropy of a system consisting of independent subsystems is equal to the sum of the entropies of the individual subsystems.
The left (right) box is initially filled with an $M$($N$)-particle gas with volume $Mv$ ($Nv$), and then the wall is removed.
For identical and different gases, Eq.~(\ref{FTAI}) gives $\langle e^{-\beta (W_{\rm id}-\Delta F_{\rm id})} \rangle_{\rm id} = 1-\lambda_{\rm id}$ and  $\langle e^{-\beta (W_{\rm dif}-\Delta F_{\rm dif})} \rangle_{\rm dif} = 1-\lambda_{\rm dif}$, respectively.
Because no work is performed upon gas mixing, $\langle e^{-\beta W_{\rm id}} \rangle_{\rm id} = \langle e^{-\beta W_{\rm dif}} \rangle_{\rm dif}=1$ holds. 
Therefore, we obtain
$\label{Flam}
	\Delta F_{\rm id} - \Delta F_{\rm dif} = \kb T \ln [{(1-\lambda_{\rm id})}/{(1-\lambda_{\rm dif})}].
$
Now, we evaluate the left-hand side of this equality.
Since $\Delta F = \Delta U -T\Delta S$ and $\Delta U_{\rm id} = \Delta U_{\rm dif}$ by the assumption of the same interaction potential, we obtain
$
	\Delta F_{\rm id} - \Delta F_{\rm dif} =
	-T(\Delta S_{\rm id} - \Delta S_{\rm dif}).
$
Moreover, since the partition functions for the two processes calculated by statistical mechanics based on ``the specific phase'' are equal, we have $\Delta S^{\rm stat}_{\rm id} = \Delta S^{\rm stat}_{\rm dif}$.
It follows then from Eq.~(\ref{Deff}) that the difference between the thermodynamic entropy productions can be evaluated only in terms of $f(N)$ as $\Delta S_{\rm id} - \Delta S_{\rm dif}=\kb[\Delta f_{\rm id}-\Delta f_{\rm dif}]$.
For identical-gas mixing, the ambiguity function of the initial state is given by $f_{\rm id}^{\rm ini}=f(M)+f(N)$ due to the assumption of additivity, whereas that of the final state is $f_{\rm id}^{\rm fin}=f(M+N)$, resulting in $\Delta f_{\rm id} = f(M+N)-f(M)-f(N)$.
In contrast, for different-gas mixing, we can quasi-statically connect the initial and final states by using a semi-permeable membrane.
As a result, the entropy difference is uniquely determined by the Clausius equality, and therefore the ambiguity function remains unchanged, and hence $\Delta f_{\rm dif}=0$.
As a consequence, we obtain
\begin{equation}\label{fAI}
	f(M+N)-f(M)-f(N) =  \ln \frac{1-\lambda_{\rm dif}}{1-\lambda_{\rm id}}.
\end{equation}
This result implies that the ambiguity in the thermodynamic entropy represented by $f(N)$ is directly related to and hence can be removed by the degree of absolute irreversibility in the gas mixing process.

To find $\lambda$, let us consider the time-reversed process of gas mixing, i.e., wall insertion as illustrated in Fig.~\ref{Fig:GM}.
After the insertion, the state may or may not return to the original one.
The events in which the state does not return to the original state are absolutely irreversible because they have no counterparts in the forward process;
the event in which the state returns to the original state is the only one without absolute irreversibility.
Therefore, $1-\lambda$ is equal to the probability that the original state is restored after the wall insertion.
Although this probability cannot be calculated explicitly in the presence of interactions, we can compare the probabilities for the two cases.
The number of combinations for the $M+N$ particles in which the left box contains $M$ particles after the wall insertion is equal to the binomial coefficient ${}_{M+N}C_M:=(M+N)!/(M!N!)$.
These partitions can be realized with equal probability due to the assumption of identical interactions.
For identical gases, all of them are identified with the original state, whereas there is only one such state for different gases.
Therefore, we obtain
$
	 1-\lambda_{\rm id}={}_{M+N}C_M(1-\lambda_{\rm dif}).
$
Then, equality~(\ref{fAI}) reduces to
$
	f(M+N)-f(M)-f(N)=-\ln{}_{M+N}C_M.
$
The solution of this functional equation is given by Eq.~(\ref{N!}).

Next, we assume Eq.~(\ref{N!}) to show Eq.~(\ref{FTAI}).
We first consider identical-gas mixing.
Let $Z(N)$ denote the partition function of the gas with the particle number $N$ and the volume $Nv$ calculated by statistical mechanics based on ``the specific phase.''
Then, from Eq.~(\ref{N!}), the thermodynamic free energy is evaluated as $F(N)=-\beta^{-1}\ln Z(N)- \beta^{-1}f(N)=-\beta^{-1}\ln [Z(N)(f(1))^N/N!]$.
Consequantly, the left-hand side of Eq.~(\ref{FTAI}) for identical-gas mixing reduces to
$
	\langle e^{-\beta(W_{\rm id}-\Delta F_{\rm id})} \rangle_{\rm id}
	=
	e^{\beta\Delta F_{\rm id}}
	=
	{}_{M+N}C_M\cdot Z(M)Z(N)/Z(M+N).
$
The right-hand side gives the probability to restore by the wall insertion the original state in which the number of particles in the left (right) partition is $M$($N$).
This is by definition equal to $1-\lambda_{\rm id}$.
Hence, Eq.~(\ref{FTAI}) for identical-gas mixing is shown.
A similar procedure vindicates Eq.~(\ref{FTAI}) for different-gas mixing.
Thus, the main result has been proven.

{\it Discussions.}
We have demonstrated that the validity of Eq.~(\ref{FTAI}) is equivalent to that of Eq.~(\ref{N!}).
Therefore, mathematically speaking, requiring either of them produces the same results.
However, from a physical point of view, thermodynamics should have an operational basis.
In this sense, it is noteworthy that Eq.~(\ref{FTAI}) defines the equilibrium free-energy difference from statistics of the work that we can operationally obtain by repeating a fixed nonequilibrium processes.
Hence, we think that requiring Eq.~(\ref{FTAI}) as the starting point is more appropriate for thermodynamics as an operational theory.

As clarified in the context of GP-I \cite{Gibbs1875,Kampen1984,Jaynes1992},
the thermodynamic entropy depends on the operational capability of the beholder.
Therefore, to deal with the thermodynamic entropy, we must define what we mean by the same and different states operationally.
This is the conventional wisdom to resolve the Gibbs paradox.
We do so by introducing different probabilities of absolute irreversibility for the two mixing processes because they are defined as the probability of time-reversed events that do not have the counterparts in the forward process.
In this sense, we follow the conventional wisdom and claim no novelty on this point.
Rather, we believe that any resolution of the Gibbs paradox is destined to follow it.

As we have discussed above the fluctuation theorem with absolute irreversibility (\ref{FTAI}) is closely related to extensivity (\ref{Ext}).
For simplicity, we consider mixing of identical ideal gases with the same temperature and volume: $(T,V,N|T,V,N)\to (T,2V,2N)$.
In the thermodynamic limit, we can explicitly show from Eq.~(\ref{FTAI}) that the thermodynamic entropy production upon gas mixing is sub-extensive, and such a sub-extensive correction is thermodynamically ignored, giving $S(T,V,N|T,V,N)=S(T,2V,2N)$.
Therefore, assuming additivity $S(T,V,N|T,V,N)=S(T,V,N)+S(T,V,N)=2S(T,V,N)$, we obtain $S(T,2V,2N)=2S(T,V,N)$, which is extensivity~(\ref{Ext}) with $q=2$.
Thus, Eq.~(\ref{FTAI}) reduces to extensivity~(\ref{Ext}) in the thermodynamic limit.

We have dealt with interacting gases by assuming that the interaction does not break additivity.
In the Pauli-Jaynes method \cite{Pauli1973,Jaynes1992}, only noninteracting gases were discussed.
Although it is possible to extend their method to interacting gases such as van der Waals gases, a general discussion seems intractable because we have to invoke the specific form of the statistical-mechanical entropy.
In contrast, our method can consider general interacting gases.
We consider both identical- and different-gas mixing and compare their entropy production to cancel out the effects of interactions.
We stress that to do so Eq.~(\ref{FTAI}) plays a crucial role.
We translate the free-energy difference $\Delta F$, which is not explicitly computable, into the absolutely irreversible probability $\lambda$, which is not either.
However, $\Delta F_{\rm id}-\Delta F_{\rm dif}$ is computable because the right-hand side of Eq.~(\ref{fAI}) can be evaluated by combinatorics.
Thus, the fluctuation theorem with absolute irreversibility~(\ref{FTAI}) is crucial to remove the ambiguity $f(N)$ in the presence of interparticle interactions, although the fluctuations of work do not exist in gas mixing.

One may wonder why we do not consider the thermodynamic entropy of an open system from the beginning.
We could extend the Clausius definition to an open system as $\Delta S=\int_{\rm q.s.} (\delta Q + \mu dN)/T$.
However, to define the thermodynamic entropy in this way, we must define the chemical potential $\mu$ beforehand.
The crucial point is that the chemical potential is defined from the very $N$-dependence of the thermodynamic entropy as $\mu=T\partial S/\partial N$.
To avoid falling into a circular argument, we cannot start from an open thermodynamic system.

Finally, we discuss why we consider isothermal systems instead of isolated systems.
First of all, the original discussion by Gibbs himself \cite{Gibbs1875} considered the isothermal setup at a constant temperature. 
Furthermore, in small isolated systems, the relation between thermodynamics and statistical mechanics is not completely understood.
On the other hand, in isothermal systems, thermodynamic responses are considered to be reproduced by statistical mechanics as long as the heat bath is infinitely large.
Thus, GP-III is more well-defined in isothermal systems than in isolated systems.

{\it Conclusion.}
We have demonstrated that the validity of the fluctuation theorem with absolute irreversibility (\ref{FTAI}) is equivalent to the specific form~(\ref{N!}) of the ambiguity function of the thermodynamic entropy in Eq.~(\ref{Deff}).
Requiring Eq.~(\ref{FTAI}) is more appropriate than requiring Eq.~(\ref{N!}) from an operational point of view.

\begin{acknowledgments}
{\it Acknowledgments.}
We appreciate useful comments by Naoyuki Sakumichi, Takashi Mori, Kiyoshi Kanazawa, Ken Funo, Yuto Ashida and Ryusuke Hamazaki.
This work was supported by
KAKENHI Grant No. JP26287088 from the Japan Society for the Promotion of Science, 
a Grant-in-Aid for Scientific Research on Innovative Areas ``Topological Materials Science (KAKENHI Grant No. JP15H05855), 
the Photon Frontier Network Program from MEXT of Japan,
and the Mitsubishi Foundation.
YM was supported by the Japan Society for the Promotion of Science through the Program for Leading Graduate Schools (MERIT) and JSPS Fellowship (JSPS KAKENHI Grant Number JP15J00410).
\end{acknowledgments}


\begin{thebibliography}{29}%
\makeatletter
\providecommand \@ifxundefined [1]{%
 \@ifx{#1\undefined}
}%
\providecommand \@ifnum [1]{%
 \ifnum #1\expandafter \@firstoftwo
 \else \expandafter \@secondoftwo
 \fi
}%
\providecommand \@ifx [1]{%
 \ifx #1\expandafter \@firstoftwo
 \else \expandafter \@secondoftwo
 \fi
}%
\providecommand \natexlab [1]{#1}%
\providecommand \enquote  [1]{``#1''}%
\providecommand \bibnamefont  [1]{#1}%
\providecommand \bibfnamefont [1]{#1}%
\providecommand \citenamefont [1]{#1}%
\providecommand \href@noop [0]{\@secondoftwo}%
\providecommand \href [0]{\begingroup \@sanitize@url \@href}%
\providecommand \@href[1]{\@@startlink{#1}\@@href}%
\providecommand \@@href[1]{\endgroup#1\@@endlink}%
\providecommand \@sanitize@url [0]{\catcode `\\12\catcode `\$12\catcode
  `\&12\catcode `\#12\catcode `\^12\catcode `\_12\catcode `\%12\relax}%
\providecommand \@@startlink[1]{}%
\providecommand \@@endlink[0]{}%
\providecommand \url  [0]{\begingroup\@sanitize@url \@url }%
\providecommand \@url [1]{\endgroup\@href {#1}{\urlprefix }}%
\providecommand \urlprefix  [0]{URL }%
\providecommand \Eprint [0]{\href }%
\providecommand \doibase [0]{http://dx.doi.org/}%
\providecommand \selectlanguage [0]{\@gobble}%
\providecommand \bibinfo  [0]{\@secondoftwo}%
\providecommand \bibfield  [0]{\@secondoftwo}%
\providecommand \translation [1]{[#1]}%
\providecommand \BibitemOpen [0]{}%
\providecommand \bibitemStop [0]{}%
\providecommand \bibitemNoStop [0]{.\EOS\space}%
\providecommand \EOS [0]{\spacefactor3000\relax}%
\providecommand \BibitemShut  [1]{\csname bibitem#1\endcsname}%
\let\auto@bib@innerbib\@empty
\bibitem [{\citenamefont {van Kampen}(1984)}]{Kampen1984}%
  \BibitemOpen
  \bibfield  {author} {\bibinfo {author} {\bibfnamefont {N.~G.}\ \bibnamefont
  {van Kampen}},\ }\href@noop {} {\emph {\bibinfo {title} {The Gibbs
  Paradox}}},\ edited by\ \bibinfo {editor} {\bibfnamefont {W.~E.}\
  \bibnamefont {Parry}}\ (\bibinfo  {publisher} {Pergamon},\ \bibinfo {year}
  {1984})\BibitemShut {NoStop}%
\bibitem [{\citenamefont {Gibbs}(5 78)}]{Gibbs1875}%
  \BibitemOpen
  \bibfield  {author} {\bibinfo {author} {\bibfnamefont {J.~W.}\ \bibnamefont
  {Gibbs}},\ }\href@noop {} {\emph {\bibinfo {title} {On the Equilibrium of
  Heterogeneous Substances}}}\ (\bibinfo  {publisher} {Connecticut Acad.
  Sci.},\ \bibinfo {year} {1875-78})\BibitemShut {NoStop}%
\bibitem [{\citenamefont {Jaynes}(1992)}]{Jaynes1992}%
  \BibitemOpen
  \bibfield  {author} {\bibinfo {author} {\bibfnamefont {E.~T.}\ \bibnamefont
  {Jaynes}},\ }\href@noop {} {\emph {\bibinfo {title} {The Gibbs Paradox}}},\
  edited by\ \bibinfo {editor} {\bibfnamefont {C.~R.}\ \bibnamefont {Smith}},
  \bibinfo {editor} {\bibfnamefont {G.~J.}\ \bibnamefont {Erickson}}, \ and\
  \bibinfo {editor} {\bibfnamefont {P.~O.}\ \bibnamefont {Neudorfer}}\
  (\bibinfo  {publisher} {Kluwer Academic Publishers},\ \bibinfo {year}
  {1992})\BibitemShut {NoStop}%
\bibitem [{\citenamefont {Ehrenfest}\ and\ \citenamefont
  {Trkal}(1921)}]{Ehrenfest1921}%
  \BibitemOpen
  \bibfield  {author} {\bibinfo {author} {\bibfnamefont {P.}~\bibnamefont
  {Ehrenfest}}\ and\ \bibinfo {author} {\bibfnamefont {V.}~\bibnamefont
  {Trkal}},\ }\href@noop {} {\bibfield  {journal} {\bibinfo  {journal} {Ann.
  Physik}\ }\textbf {\bibinfo {volume} {65}},\ \bibinfo {pages} {609} (\bibinfo
  {year} {1921})}\BibitemShut {NoStop}%
\bibitem [{\citenamefont {Gibbs}(1902)}]{Gibbs1902}%
  \BibitemOpen
  \bibfield  {author} {\bibinfo {author} {\bibfnamefont {J.~W.}\ \bibnamefont
  {Gibbs}},\ }\href@noop {} {\emph {\bibinfo {title} {Elementary Principles in
  Statistical Mehcnics}}}\ (\bibinfo  {publisher} {Yale University Press},\
  \bibinfo {year} {1902})\BibitemShut {NoStop}%
\bibitem [{\citenamefont {Sekimoto}(2010)}]{SekimotoBook}%
  \BibitemOpen
  \bibfield  {author} {\bibinfo {author} {\bibfnamefont {K.}~\bibnamefont
  {Sekimoto}},\ }\href@noop {} {\emph {\bibinfo {title} {Stochastic
  Enegetics}}}\ (\bibinfo  {publisher} {Springer},\ \bibinfo {year}
  {2010})\BibitemShut {NoStop}%
\bibitem [{\citenamefont {Klages}\ \emph {et~al.}(2013)\citenamefont {Klages},
  \citenamefont {Just},\ and\ \citenamefont {Jarzynski}}]{JarzynskiBook}%
  \BibitemOpen
  \bibinfo {editor} {\bibfnamefont {R.}~\bibnamefont {Klages}}, \bibinfo
  {editor} {\bibfnamefont {W.}~\bibnamefont {Just}}, \ and\ \bibinfo {editor}
  {\bibfnamefont {C.}~\bibnamefont {Jarzynski}},\ eds.,\ \href@noop {} {\emph
  {\bibinfo {title} {Nonequilibrium Statistical Physics of Small Systems}}}\
  (\bibinfo  {publisher} {Wiley-VCH},\ \bibinfo {year} {2013})\BibitemShut
  {NoStop}%
\bibitem [{\citenamefont {Seifert}(2012)}]{SeifertRev}%
  \BibitemOpen
  \bibfield  {author} {\bibinfo {author} {\bibfnamefont {U.}~\bibnamefont
  {Seifert}},\ }\href@noop {} {\bibfield  {journal} {\bibinfo  {journal} {Rep.
  Prog. Phys.}\ }\textbf {\bibinfo {volume} {75}},\ \bibinfo {pages} {126001}
  (\bibinfo {year} {2012})}\BibitemShut {NoStop}%
\bibitem [{\citenamefont {Pauli}(1973)}]{Pauli1973}%
  \BibitemOpen
  \bibfield  {author} {\bibinfo {author} {\bibfnamefont {W.}~\bibnamefont
  {Pauli}},\ }\href@noop {} {\emph {\bibinfo {title} {Thermodynamics and the
  Kinetic Theory of Gases}}}\ (\bibinfo  {publisher} {MIT Press},\ \bibinfo
  {year} {1973})\BibitemShut {NoStop}%
\bibitem [{\citenamefont {Evans}\ \emph {et~al.}(1993)\citenamefont {Evans},
  \citenamefont {Cohen},\ and\ \citenamefont
  {Morriss}}]{EvansCohenMorriss1993}%
  \BibitemOpen
  \bibfield  {author} {\bibinfo {author} {\bibfnamefont {D.~J.}\ \bibnamefont
  {Evans}}, \bibinfo {author} {\bibfnamefont {E.~G.~D.}\ \bibnamefont {Cohen}},
  \ and\ \bibinfo {author} {\bibfnamefont {G.~P.}\ \bibnamefont {Morriss}},\
  }\href@noop {} {\bibfield  {journal} {\bibinfo  {journal} {Phys. Rev. Lett.}\
  }\textbf {\bibinfo {volume} {71}},\ \bibinfo {pages} {2401} (\bibinfo {year}
  {1993})}\BibitemShut {NoStop}%
\bibitem [{\citenamefont {Gallavotti}\ and\ \citenamefont
  {Cohen}(1995)}]{GallavottiCohen1995PRL}%
  \BibitemOpen
  \bibfield  {author} {\bibinfo {author} {\bibfnamefont {G.}~\bibnamefont
  {Gallavotti}}\ and\ \bibinfo {author} {\bibfnamefont {E.~G.~D.}\ \bibnamefont
  {Cohen}},\ }\href@noop {} {\bibfield  {journal} {\bibinfo  {journal} {Phys.
  Rev. Lett.}\ }\textbf {\bibinfo {volume} {74}},\ \bibinfo {pages} {2694}
  (\bibinfo {year} {1995})}\BibitemShut {NoStop}%
\bibitem [{\citenamefont {Jarzynski}(1997)}]{Jarzynski1997PRL}%
  \BibitemOpen
  \bibfield  {author} {\bibinfo {author} {\bibfnamefont {C.}~\bibnamefont
  {Jarzynski}},\ }\href@noop {} {\bibfield  {journal} {\bibinfo  {journal}
  {Phys. Rev. Lett.}\ }\textbf {\bibinfo {volume} {78}},\ \bibinfo {pages}
  {2690} (\bibinfo {year} {1997})}\BibitemShut {NoStop}%
\bibitem [{\citenamefont {Crooks}(1999)}]{Crooks1999}%
  \BibitemOpen
  \bibfield  {author} {\bibinfo {author} {\bibfnamefont {G.~E.}\ \bibnamefont
  {Crooks}},\ }\href@noop {} {\bibfield  {journal} {\bibinfo  {journal} {Phys.
  Rev. E}\ }\textbf {\bibinfo {volume} {60}},\ \bibinfo {pages} {2721}
  (\bibinfo {year} {1999})}\BibitemShut {NoStop}%
\bibitem [{\citenamefont {Hatano}\ and\ \citenamefont
  {Sasa}(2001)}]{HatanoSasa2001}%
  \BibitemOpen
  \bibfield  {author} {\bibinfo {author} {\bibfnamefont {T.}~\bibnamefont
  {Hatano}}\ and\ \bibinfo {author} {\bibfnamefont {S. I.}\ \bibnamefont
  {Sasa}},\ }\href@noop {} {\bibfield  {journal} {\bibinfo  {journal} {Phys.
  Rev. Lett.}\ }\textbf {\bibinfo {volume} {86}},\ \bibinfo {pages} {3463}
  (\bibinfo {year} {2001})}\BibitemShut {NoStop}%
\bibitem [{\citenamefont {Seifert}(2005)}]{Seifert2005}%
  \BibitemOpen
  \bibfield  {author} {\bibinfo {author} {\bibfnamefont {U.}~\bibnamefont
  {Seifert}},\ }\href@noop {} {\bibfield  {journal} {\bibinfo  {journal} {Phys.
  Rev. Lett.}\ }\textbf {\bibinfo {volume} {95}},\ \bibinfo {pages} {040602}
  (\bibinfo {year} {2005})}\BibitemShut {NoStop}%
\bibitem [{\citenamefont {Jarzynski}(2011)}]{JarzynskiRev}%
  \BibitemOpen
  \bibfield  {author} {\bibinfo {author} {\bibfnamefont {C.}~\bibnamefont
  {Jarzynski}},\ }\href@noop {} {\bibfield  {journal} {\bibinfo  {journal}
  {Annual Review of Condensed Matter Physics}\ }\textbf {\bibinfo {volume}
  {2}},\ \bibinfo {pages} {329} (\bibinfo {year} {2011})}\BibitemShut {NoStop}%
\bibitem [{\citenamefont {Van~den Broeck}\ and\ \citenamefont
  {Esposito}(2015)}]{BroeckEspositoRev}%
  \BibitemOpen
  \bibfield  {author} {\bibinfo {author} {\bibfnamefont {C.}~\bibnamefont
  {Van~den Broeck}}\ and\ \bibinfo {author} {\bibfnamefont {M.}~\bibnamefont
  {Esposito}},\ }\href {\doibase 10.1016/j.physa.2014.04.035} {\bibfield
  {journal} {\bibinfo  {journal} {Physica A}\ }\textbf {\bibinfo {volume}
  {418}},\ \bibinfo {pages} {6} (\bibinfo {year} {2015})}\BibitemShut {NoStop}%
\bibitem [{\citenamefont {Murashita}\ \emph {et~al.}(2014)\citenamefont
  {Murashita}, \citenamefont {Funo},\ and\ \citenamefont
  {Ueda}}]{MurashitaFunoUeda2014}%
  \BibitemOpen
  \bibfield  {author} {\bibinfo {author} {\bibfnamefont {Y.}~\bibnamefont
  {Murashita}}, \bibinfo {author} {\bibfnamefont {K.}~\bibnamefont {Funo}}, \
  and\ \bibinfo {author} {\bibfnamefont {M.}~\bibnamefont {Ueda}},\ }\href@noop
  {} {\bibfield  {journal} {\bibinfo  {journal} {Phys. Rev. E}\ }\textbf
  {\bibinfo {volume} {90}},\ \bibinfo {pages} {042110} (\bibinfo {year}
  {2014})}\BibitemShut {NoStop}%
\bibitem [{\citenamefont {Planck}(1925)}]{Planck1925}%
  \BibitemOpen
  \bibfield  {author} {\bibinfo {author} {\bibfnamefont {M.}~\bibnamefont
  {Planck}},\ }\href@noop {} {\bibfield  {journal} {\bibinfo  {journal} {S.-B.
  Preu{\ss}. Akad. Wiss.}\ ,\ \bibinfo {pages} {49}} (\bibinfo {year}
  {1925})}\BibitemShut {NoStop}%
\bibitem [{\citenamefont {Schr\"odinger}(1946)}]{Schrodinger1946}%
  \BibitemOpen
  \bibfield  {author} {\bibinfo {author} {\bibfnamefont {E.}~\bibnamefont
  {Schr\"odinger}},\ }\href@noop {} {\emph {\bibinfo {title} {Statistical
  Thermodynamics}}}\ (\bibinfo  {publisher} {Cambridge University Press},\
  \bibinfo {year} {1946})\BibitemShut {NoStop}%
\bibitem [{\citenamefont {Feynman}(1972)}]{Feynman1972}%
  \BibitemOpen
  \bibfield  {author} {\bibinfo {author} {\bibfnamefont {R.~P.}\ \bibnamefont
  {Feynman}},\ }\href@noop {} {\emph {\bibinfo {title} {Statistical Mechanics:
  A Set of Lectures}}}\ (\bibinfo  {publisher} {Westview Press},\ \bibinfo
  {year} {1972})\BibitemShut {NoStop}%
\bibitem [{\citenamefont {Zubarev}(1974)}]{Zubarev1974}%
  \BibitemOpen
  \bibfield  {author} {\bibinfo {author} {\bibfnamefont {D.}~\bibnamefont
  {Zubarev}},\ }\href@noop {} {\emph {\bibinfo {title} {Nonequilibrium
  Statistical Thermodynamics}}}\ (\bibinfo  {publisher} {Plenum Pub. Corp.},\
  \bibinfo {year} {1974})\BibitemShut {NoStop}%
\bibitem [{\citenamefont {Landau}\ and\ \citenamefont
  {Lifshitz}(1975)}]{LandauLifshitz1975}%
  \BibitemOpen
  \bibfield  {author} {\bibinfo {author} {\bibfnamefont {L.~D.}\ \bibnamefont
  {Landau}}\ and\ \bibinfo {author} {\bibfnamefont {E.~M.}\ \bibnamefont
  {Lifshitz}},\ }\href@noop {} {\emph {\bibinfo {title} {Statistical
  Physics}}}\ (\bibinfo  {publisher} {Butterworth-Heinemann},\ \bibinfo {year}
  {1975})\BibitemShut {NoStop}%
\bibitem [{\citenamefont {Toda}\ \emph {et~al.}(1978)\citenamefont {Toda},
  \citenamefont {Kubo},\ and\ \citenamefont {Saito}}]{Kubo1978}%
  \BibitemOpen
  \bibfield  {author} {\bibinfo {author} {\bibfnamefont {M.}~\bibnamefont
  {Toda}}, \bibinfo {author} {\bibfnamefont {R.}~\bibnamefont {Kubo}}, \ and\
  \bibinfo {author} {\bibfnamefont {N.}~\bibnamefont {Saito}},\ }\href@noop {}
  {\emph {\bibinfo {title} {Statistical Physics I: Equilibrium Statistical
  Mechanics}}}\ (\bibinfo  {publisher} {Springer},\ \bibinfo {year}
  {1978})\BibitemShut {NoStop}%
\bibitem [{\citenamefont {Callen}(1985)}]{Callen1985}%
  \BibitemOpen
  \bibfield  {author} {\bibinfo {author} {\bibfnamefont {H.~B.}\ \bibnamefont
  {Callen}},\ }\href@noop {} {\emph {\bibinfo {title} {Thermodynamics and an
  Introduction to Thermostatistics}}}\ (\bibinfo  {publisher} {John Wiley \&
  Sons},\ \bibinfo {year} {1985})\BibitemShut {NoStop}%
\bibitem [{\citenamefont {Kondepudi}\ and\ \citenamefont
  {Prigogine}(1998)}]{Prigogine1998}%
  \BibitemOpen
  \bibfield  {author} {\bibinfo {author} {\bibfnamefont {D.}~\bibnamefont
  {Kondepudi}}\ and\ \bibinfo {author} {\bibfnamefont {I.}~\bibnamefont
  {Prigogine}},\ }\href@noop {} {\emph {\bibinfo {title} {Modern
  Thermodynamics: From Heat Engines to Dissipative Structures}}}\ (\bibinfo
  {publisher} {John Wiley \& Sons},\ \bibinfo {year} {1998})\BibitemShut
  {NoStop}%
\bibitem [{\citenamefont {Frenkel}(2014)}]{Frenkel2014}%
  \BibitemOpen
  \bibfield  {author} {\bibinfo {author} {\bibfnamefont {D.}~\bibnamefont
  {Frenkel}},\ }\href@noop {} {\bibfield  {journal} {\bibinfo  {journal} {Mol.
  Phys.}\ }\textbf {\bibinfo {volume} {112}},\ \bibinfo {pages} {2325}
  (\bibinfo {year} {2014})}\BibitemShut {NoStop}%
\bibitem [{\citenamefont {Cates}\ and\ \citenamefont
  {Manoharan}(2015)}]{CatesManoharan2015}%
  \BibitemOpen
  \bibfield  {author} {\bibinfo {author} {\bibfnamefont {M.~E.}\ \bibnamefont
  {Cates}}\ and\ \bibinfo {author} {\bibfnamefont {V.~N.}\ \bibnamefont
  {Manoharan}},\ }\href@noop {} {\bibfield  {journal} {\bibinfo  {journal}
  {Soft Matter}\ }\textbf {\bibinfo {volume} {11}},\ \bibinfo {pages} {6538}
  (\bibinfo {year} {2015})}\BibitemShut {NoStop}%
\bibitem [{\citenamefont {Ashida}\ \emph {et~al.}(2014)\citenamefont {Ashida},
  \citenamefont {Funo}, \citenamefont {Murashita},\ and\ \citenamefont
  {Ueda}}]{AshidaFunoMurashitaUeda2014}%
  \BibitemOpen
  \bibfield  {author} {\bibinfo {author} {\bibfnamefont {Y.}~\bibnamefont
  {Ashida}}, \bibinfo {author} {\bibfnamefont {K.}~\bibnamefont {Funo}},
  \bibinfo {author} {\bibfnamefont {Y.}~\bibnamefont {Murashita}}, \ and\
  \bibinfo {author} {\bibfnamefont {M.}~\bibnamefont {Ueda}},\ }\href@noop {}
  {\bibfield  {journal} {\bibinfo  {journal} {Phys. Rev. E}\ }\textbf {\bibinfo
  {volume} {90}},\ \bibinfo {pages} {052125} (\bibinfo {year}
  {2014})}\BibitemShut {NoStop}%
\end{thebibliography}

%

\end{document}